\documentclass[12pt]{article}

\usepackage{amsmath}
\usepackage{stmaryrd}
\usepackage{subfigure}
\usepackage{empheq}
\usepackage{appendix,epic,eepic,amsmath,amsthm,amssymb,epsfig,cite,indentfirst,oldgerm}

\newcommand{\EQ}{\begin{equation}}
\newcommand{\EN}{\end{equation}}

\newcommand{\bear}{\begin{eqnarray}}
\newcommand{\ear}{\end{eqnarray}}
\newcommand{\bt} { \begin{tabular} }
\newcommand{\et}{ \end{tabular} }
\newcommand{\bc} { \begin{center} }
\newcommand{\ec}{ \end{center} }

\newcommand{\btb} { \begin{table} }
\newcommand{\etb}{ \end{table} }

\begin{document}

\topmargin 0pt
\oddsidemargin 5mm
\newcommand{\NP}[1]{Nucl.\ Phys.\ {\bf #1}}
\newcommand{\PL}[1]{Phys.\ Lett.\ {\bf #1}}
\newcommand{\NC}[1]{Nuovo Cimento {\bf #1}}
\newcommand{\CMP}[1]{Comm.\ Math.\ Phys.\ {\bf #1}}
\newcommand{\PR}[1]{Phys.\ Rev.\ {\bf #1}}
\newcommand{\PRL}[1]{Phys.\ Rev.\ Lett.\ {\bf #1}}
\newcommand{\MPL}[1]{Mod.\ Phys.\ Lett.\ {\bf #1}}
\newcommand{\JETP}[1]{Sov.\ Phys.\ JETP {\bf #1}}
\newcommand{\TMP}[1]{Teor.\ Mat.\ Fiz.\ {\bf #1}}

\renewcommand{\thefootnote}{\fnsymbol{footnote}}

\newpage
\setcounter{page}{0}
\begin{titlepage}
\begin{flushright}

\end{flushright}
\vspace{0.5cm}
\begin{center}
{\large The symmetric six-vertex model and the Segre cubic threefold} \\
\vspace{1cm}
{\large M.J. Martins } \\
\vspace{0.15cm}
{\em Universidade Federal de S\~ao Carlos\\
Departamento de F\'{\i}sica \\
C.P. 676, 13565-905, S\~ao Carlos (SP), Brazil\\}
\vspace{0.35cm}
\end{center}
\vspace{0.5cm}
\vspace{0.5cm}
\begin{abstract}
In this paper we investigate the mathematical properties of the integrability
of the symmetric six-vertex model towards the view of Algebraic Geometry.
We show that the algebraic variety originated from 
Baxter's commuting transfer method is 
birationally isomorphic to a 
ubiquitous threefold known as
Segre cubic primal. This relation makes it possible
to present the most generic
solution for the Yang-Baxter triple associated 
to this lattice model.
The respective $\mathrm{R}$-matrix
and Lax operators are 
parametrized by three independent affine spectral variables.
\end{abstract}

\vspace{.15cm} \centerline{}
\vspace{.1cm} \centerline{Keywords: Integrable models, Algebraic Geometry}
\vspace{.15cm} \centerline{May~~2015}

\end{titlepage}


\pagestyle{empty}

\newpage

\pagestyle{plain}
\pagenumbering{arabic}

\renewcommand{\thefootnote}{\arabic{footnote}}
\newtheorem{proposition}{Proposition}
\newtheorem{pr}{Proposition}
\newtheorem{remark}{Remark}
\newtheorem{re}{Remark}
\newtheorem{theorem}{Theorem}
\newtheorem{theo}{Theorem}

\def\ll{\left\lgroup}
\def\rr{\right\rgroup}

\newtheorem{Theorem}{Theorem}[section]
\newtheorem{Corollary}[Theorem]{Corollary}
\newtheorem{Proposition}[Theorem]{Proposition}
\newtheorem{Conjecture}[Theorem]{Conjecture}
\newtheorem{Lemma}[Theorem]{Lemma}
\newtheorem{Example}[Theorem]{Example}
\newtheorem{Note}[Theorem]{Note}
\newtheorem{Definition}[Theorem]{Definition}

\section{Introduction}

The six-vertex model is a system of statistical
mechanics defined on a square
$\mathrm{N} \times \mathrm{N}$ lattice in which 
the configurations  sit
on the bonds connecting
neighboring lattice points \cite{LW}.
Its origin
goes back to a model introduced by Pauling to explain
the experimental fact that the ice crystal 
has a residual entropy 
at very low temperatures \cite{PAU}. The lattice sites
are assumed to be occupied by oxygen ions having four nearest hydrogen
neighbors. The hydrogen distribution is such that two
of them are close to the oxygen ion and the other two  
are farther away. This means that the sixteen possible
states at each lattice vertex are reduced to only six 
which are shown in Figure \ref{figure1}.
\setlength{\unitlength}{2500sp}
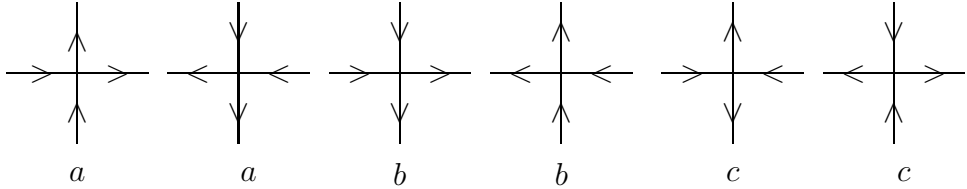
\begin{figure}[ht]
\begin{center}
\begin{picture}(8000,2000)
{\put(-700,900){\line(1,0){1400}}}
{\put(900,900){\line(1,0){1400}}}
{\put(2500,900){\line(1,0){1400}}}
{\put(4100,900){\line(1,0){1400}}}
{\put(5800,900){\line(1,0){1400}}}
{\put(7400,900){\line(1,0){1400}}}
{\put(0,1600){\line(0,-1){1400}}}
{\put(1600,1600){\line(0,-1){1400}}}
{\put(3200,1600){\line(0,-1){1400}}}
{\put(4800,1600){\line(0,-1){1400}}}
{\put(6500,1600){\line(0,-1){1400}}}
{\put(8100,1600){\line(0,-1){1400}}}
{\put(-350,890){\makebox(0,0){\fontsize{12}{14}\selectfont $>$}}}
{\put(1200,890){\makebox(0,0){\fontsize{12}{14}\selectfont $<$}}}
{\put(2800,890){\makebox(0,0){\fontsize{12}{14}\selectfont $>$}}}
{\put(4400,890){\makebox(0,0){\fontsize{12}{14}\selectfont $<$}}}
{\put(6100,890){\makebox(0,0){\fontsize{12}{14}\selectfont $>$}}}
{\put(7700,890){\makebox(0,0){\fontsize{12}{14}\selectfont $<$}}}
{\put(400,890){\makebox(0,0){\fontsize{12}{14}\selectfont $>$}}}
{\put(2000,890){\makebox(0,0){\fontsize{12}{14}\selectfont $<$}}}
{\put(3600,890){\makebox(0,0){\fontsize{12}{14}\selectfont $>$}}}
{\put(5200,890){\makebox(0,0){\fontsize{12}{14}\selectfont $<$}}}
{\put(6900,890){\makebox(0,0){\fontsize{12}{14}\selectfont $<$}}}
{\put(8500,890){\makebox(0,0){\fontsize{12}{14}\selectfont $>$}}}
{\put(0,1200){\makebox(0,0){\fontsize{12}{14}\selectfont $\wedge$}}}
{\put(1600,1300){\makebox(0,0){\fontsize{12}{14}\selectfont $\vee$}}}
{\put(3200,1300){\makebox(0,0){\fontsize{12}{14}\selectfont $\vee$}}}
{\put(4800,1300){\makebox(0,0){\fontsize{12}{14}\selectfont $\wedge$}}}
{\put(6500,1300){\makebox(0,0){\fontsize{12}{14}\selectfont $\wedge$}}}
{\put(8100,1300){\makebox(0,0){\fontsize{12}{14}\selectfont $\vee$}}}
{\put(0,500){\makebox(0,0){\fontsize{12}{14}\selectfont $\wedge$}}}
{\put(1600,500){\makebox(0,0){\fontsize{12}{14}\selectfont $\vee$}}}
{\put(3200,500){\makebox(0,0){\fontsize{12}{14}\selectfont $\vee$}}}
{\put(4800,500){\makebox(0,0){\fontsize{12}{14}\selectfont $\wedge$}}}
{\put(6500,500){\makebox(0,0){\fontsize{12}{14}\selectfont $\vee$}}}
{\put(8100,500){\makebox(0,0){\fontsize{12}{14}\selectfont $\wedge$}}}
{\put(0,-100){\makebox(0,0){\fontsize{12}{14}\selectfont $a$}}}
{\put(1700,-100){\makebox(0,0){\fontsize{12}{14}\selectfont $a$}}}
{\put(3200,-100){\makebox(0,0){\fontsize{12}{14}\selectfont $b$}}}
{\put(4800,-100){\makebox(0,0){\fontsize{12}{14}\selectfont $b$}}}
{\put(6500,-100){\makebox(0,0){\fontsize{12}{14}\selectfont $c$}}}
{\put(8200,-100){\makebox(0,0){\fontsize{12}{14}\selectfont $c$}}}
\end{picture}
\end{center}
\caption{The allowed configurations of the symmetric 
six-vertex model with
the respective energy weights $a,b$ and $c$.}
\label{figure1}
\end{figure}

The arrows tip pointing inwards represent states where
the hydrogens are close to the oxygen while those outwards
represent the hydrogens in more distant positions. In the 
symmetric case the energy interactions
are unchanged by reversing the arrows given rise to only three
distinct weights $a,b$ and $c$. It turns out that
the model mathematical structure can be captured 
in terms of a single local operator usually denominated
transition or Lax operator. This operator acts on the 
direct product of the spaces built out of 
the horizontal and vertical degrees of freedom and its 
elements are constituted by the energy weights. Since
we have two states
per bond the Lax operator can be viewed as
a two-dimensional matrix whose elements 
are operators acting on the space
$\mathcal{H}=\mathbb{C}^{2\otimes \mathrm{N}}$ \cite{LW,FAD}.
Its expression in a suitable
basis order
can be written as,
\EQ
\label{lax}
\mathrm{L}_j(a,b,c)=\left(
\begin{array}{cc}
ae_{11}^{(j)} + be_{22}^{(j)} & c e_{21}^{(j)} \\
c e_{12}^{(j)} &  be_{11}^{(j)} + ae_{22}^{(j)}\\
\end{array}
\right),~~j=1,\cdots,\mathrm{N},
\EN
where $e_{lk}^{(j)}$ denote $2\times 2$ Weyl matrices acting 
on the subspace $\mathbb{C}^2_j \in \mathcal{H}$.

In equilibrium statistical mechanics the macroscopic quantities such
as the free energy per vertex can be computed 
after determining the sum 
$\mathcal{Z}_{\mathrm{N}}$ of the probability distribution. 
One way to calculate such partition
function is to arrange the summations
in separated sums over rows of bond state variables. The sum of a single 
lattice row defines an operator
called the model transfer matrix. Assuming periodic 
boundary conditions 
the transfer matrix
can be written by means of an ordered product 
of Lax operator as follows \cite{LW,FAD},
\EQ
\mathrm{T}(a,b,c) = \mathrm{Tr}_{2}\left[
\mathrm{L}_{\mathrm{N}}(a,b,c)
\mathrm{L}_{\mathrm{N}-1}(a,b,c)\cdots
\mathrm{L}_1(a,b,c)\right],
\EN
where the multiplications and the trace operation
are performed on the two-dimensional matrix space. 
We see that the transfer matrix acts on the space $\mathcal{H}$
and in terms of this operator the partition function is given by,
\EQ
\mathcal{Z}_{\mathrm{N}}= \mathrm{Tr}_{\mathcal{H}}[\mathrm{T}(a,b,c)^{\mathrm{N}}].
\EN

As pointed out by Baxter \cite{BAX} an important ingredient 
to determine the dominant
large $\mathrm{N}$ behaviour of the partition function is 
the existence of a family
of commuting transfer matrices. This requirement implies the condition, 
\EQ
\label{COM}
[\mathrm{T}(a^{'},b^{'},c^{'}),\mathrm{T}(a^{''},b^{''},c^{''})]=0,
\EN
for different set of weights
$a^{'},b^{'},c^{'}$ and $a^{''},b^{''},c^{''}$.

In addition, Baxter argued that a sufficient condition 
for the above commutation to hold is the
existence of an invertible $4 \times 4$ matrix $\check{\mathrm{R}}$ 
satisfying the celebrated Yang-Baxter equation \cite{BAX},
\EQ
\label{eqYB}
\check{\mathrm{R}}
\mathrm{L}_{j}(a^{'},b^{'},c^{'})
\otimes \mathrm{L}_{j}(a^{''},b^{''},c^{''})=
\mathrm{L}_{j}(a^{''},b^{''},c^{''})
\otimes \mathrm{L}_{j}(a^{'},b^{'},c^{'})
\check{\mathrm{R}},
\EN
where the tensor product is taken with respect to 
the two-dimensional space of the transition operator.

The Yang-Baxter equation consists of a set of 
algebraic relations on the energy weights and the 
$\check{\mathrm{R}}$-matrix entries. By eliminating 
the latter elements one finds the conditions on
the Lax operator weights that permit a non-trivial
solution to Eq.(\ref{eqYB}). This approach has 
been applied for
the symmetrical six-vertex model \cite{BAX}
and was found that one such solution 
exists provided the weights are the following subset
of the product of two projective spaces,
\EQ
\label{three}
\mathrm{X}=\{ (a^{'}:b^{'}:c^{'}) \times 
(a^{''}:b^{''}:c^{''}) \in \mathbb{CP}^2 \times \mathbb{CP}^2 | \mathrm{F}(
(a^{'},b^{'},c^{'}; 
a^{''},b^{''}c^{''})=0 \},
\EN
where the expression of the respective bihomogenous polynomial is,
\EQ
\label{POLBI}
\mathrm{F}
(a^{'},b^{'},c^{'}; 
a^{''},b^{''}c^{''})=
\left[(a^{'})^2+(b^{'})^2-(c^{'})^2\right]a^{''}b^{''}-
\left[(a^{''})^2+(b^{''})^2-(c^{''})^2\right]a^{'}b^{'}. 
\EN

It is not difficult to see that the codimension 
of the variety $\mathrm{X}$ 
is one and consequently we are dealing with 
a three-dimensional algebraic manifold. This threefold
contains a two-dimensional subvariety 
in which the single
primed variables are separable from the double
primed ones. This particular divisor is formally
defined as the product of two identical
algebraic sets
\footnote{Note
that the intersection
multiplicity of the divisor $\mathrm{Y}$ at $\mathrm{X}$ is one.},
\EQ
\label{DIV}
\mathrm{Y}=\{ (a^{'}:b^{'}:c^{'}) \times 
(a^{''}:b^{''}:c^{''}) \in \mathbb{CP}^2 \times \mathbb{CP}^2 | \mathrm{D}
(a^{'},b^{'},c^{'})=0; 
\mathrm{D}(a^{''},b^{''}c^{''})=0 \},
\EN
where the respective polynomial is the quadric,
\EQ
\label{QUA}
\mathrm{D}(a,b,c)=a^2+b^2-c^2-\Delta ab,~~\Delta \in \mathbb{C}.
\EN

The above divisor has the appealing feature of providing
two independent  Lax operators
sited on the same algebraic variety. The commutativity
of the respective transfer matrices means that 
their eigenvectors depend on the
energy weights only through the parameter $\Delta$. 
This property has been important for the exact solution 
of the six-vertex model by Bethe 
ansatz methods \cite{LW,FAD}.

However, the divisor $\mathrm{Y}$ does not provide us
the most arbitrary
solution to the Yang-Baxter triple (\ref{eqYB}) 
since it is just
a subset of $\mathrm{X}$. The understanding of
such general solution needs the comprehension of the geometric
properties of the original threefold $\mathrm{X}$ which is
the aim of the present work.
Interesting enough, we shall
show that the variety $\mathrm{X}$ is birational 
to a rather distinguished
cubic threefold devised long ago
by Segre \cite{SEG}. In fact, this is the unique cubic threefold having
the maximum possible number of ordinary
double points as singularities. This connection makes it possible
to present the generic solution  for the
Yang-Baxter triple parametrized in terms of three 
distinct affine spectral coordinates.

We have organized this paper as follows. We start next 
section reviewing
the Baxter approach for the symmetrical six-vertex model providing
the means to state our result for the generic solution  
of the Yang-Baxter triple.  In section \ref{sec3} we elaborate on 
the demonstration that $\mathrm{X}$ is birationally equivalent to
the Segre cubic primal. As byproduct we obtain the parameterization
of the single and double prime energy weights.

\section{The Yang-Baxter and main results}

We recall that the Yang-Baxter equation for the six vertex model can be rewritten as a 
relation acting on the tensor product 
of three spaces $\mathbb{C}^2$. Defining $\check{\mathrm{R}}=\mathrm{P}\mathrm{R}$ where
$\mathrm{P}$ denotes the transposition on $\mathbb{C}^2 \otimes \mathbb{C}^2$
one finds,
\EQ
\label{eqYB1}
\mathrm{R}_{12}
\mathrm{L}_{13}(a^{'},b^{'},c^{'})
\mathrm{L}_{23}(a^{''},b^{''},c^{''})=
\mathrm{L}_{23}(a^{''},b^{''},c^{''})
\mathrm{L}_{13}(a^{'},b^{'},c^{'})
\mathrm{R}_{12},
\EN

The operator $\hat{\mathrm{O}}_{ij} \in \mathrm{C}^2 \otimes \mathrm{C}^2 \otimes \mathrm{C}^2$
acts as $4 \times 4$ matrix $\hat{\mathrm{O}}$ on the $i\mathrm{th}$ and
the $j\mathrm{th}$ subspaces and as identity on the remaining tensor space component.
From Eq.(\ref{lax}) we obtain the matrix representation of the transition operator,
\EQ
\label{lax1}
\mathrm{L}(a,b,c)=\left(
\begin{array}{cccc}
a & 0 & 0 & 0 \\
0 & b & c & 0 \\
0 & c & b & 0 \\
0 & 0 & 0 & a \\
\end{array}
\right).
\EN

Considering the above Lax operator structure we 
take an arbitrary $\mathrm{R}$-matrix and substitute both 
of them on the Yang-Baxter
equation (\ref{eqYB1}). These functional relations under 
the assumption of non-null energy weights 
fix many elements of the $\mathrm{R}$-matrix to be zero. 
As a result it is sufficient to choose 
the $\mathrm{R}$-matrix with the same 
form of the Lax operator,
\EQ
\label{RMA}
\mathrm{R}=\left(
\begin{array}{cccc}
\bf{a} & 0 & 0 & 0 \\
0 & \bf{b} & \bf{c} & 0 \\
0 & \bf{c} & \bf{b} & 0 \\
0 & 0 & 0 & \bf{a} \\
\end{array}
\right).
\EN
where bold letters are used to distinguish 
$\mathrm{R}$-matrix entries of the energy weights.

Considering the structure of the Lax operator and
the $\mathrm{R}$-matrix one finds that the respective
Yang-Baxter equation (\ref{eqYB1}) reduces to only
three distinct relations \cite{BAX},
\begin{eqnarray}
\label{threeq1}
&& {\bf{a}} c^{'} a^{''}-{\bf{b}}c^{'}b^{''}-{\bf{c}}a^{'}c^{''}=0,\\
\label{threeq2}
&& {\bf{c}} b^{'} a^{''}-{\bf{c}}a^{'}b^{''}-{\bf{b}}c^{'}c^{''}=0,\\
\label{threeq3}
&& {\bf{c}} c^{'} b^{''}+{\bf{b}}a^{'}c^{''}-{\bf{a}}b^{'}c^{''}=0.
\end{eqnarray}

The vanishing of the above polynomials 
defines an algebraic variety
now on the product of three 
projective spaces 
$\mathbb{CP}^2 \times \mathbb{CP}^2 \times \mathbb{CP}^2$.
This algebraic set is strictly speaking reducible since
its ideal can be rewritten as the intersection of five 
distinct primary ideals. This conclusion was reached by
studying the ideal primary decomposition with the help 
of the computer algebra package Singular \cite{SIN}. 
However, here we are interested in genuine six-vertex
systems in which none of the energy weights vanish 
everywhere. For instance, we have to disregard  
the subset of zeros which are sited at the 
lines $c^{'}=0$ or $c^{''}=0$ of the projective
spaces associated to the Lax operators. It turns out that such restriction
is satisfied by only one of the five 
possible irreducible components. 
This is equivalent 
to the condition that the 
system of homogeneous relations (\ref{threeq1}-\ref{threeq2})
must have non trivial solution for the unknowns ${\bf{a}},{\bf{b}}$
and ${\bf{c}}$, namely
\EQ
\left|
\begin{array}{ccc}
c^{'} a^{''} & -c^{'} b^{''} & -a^{'}c^{''} \\
0& -c^{'} c^{''} & b^{'}a^{''}-a^{'}b^{''} \\
-b^{'} c^{''} & a^{'} c^{''} & c^{'}b^{''} \\
\end{array}
\right|=0.
\EN

This determinant becomes proportional to the bihomogenous polynomial (\ref{POLBI}) defining
the threefold $\mathrm{X}$.  
The polynomials (\ref{threeq1}-\ref{threeq2}) can now be solved for the 
ratios ${\bf{a}}:{\bf{b}}:{\bf{c}}$ and as a result we obtain,
\EQ
\label{ratios}
\frac{{\bf{a}}}{{\bf{c}}}=\frac{(b^{'}a^{''}-a^{'}b^{''})b^{''}}{c^{'}a^{''}c^{''}}+\frac{a^{'}c^{''}}{c^{'}a^{''}},~~
\frac{{\bf{b}}}{{\bf{c}}}=\frac{b^{'}a^{''}-a^{'}b^{''}}{c^{'}c^{''}}.
\EN

At this point we have introduced the basic ingredients 
to state our final formulae
for the generic solution to the Yang-Baxter triple. Let us
denote by the $\mu_1,\mu_2$ and $\mu_3$ the affine 
spectral variables
used to parameterize 
the algebraic threefold $\mathrm{X}$. We find that 
Eq.(\ref{eqYB1}) can be rewritten as follows, 
\EQ
\label{eqYB2}
\mathrm{R}_{12}(\mu_1,\mu_2,\mu_3)
\mathrm{L}_{13}(\mu_1,\mu_2,\mu_3)
\mathrm{L}_{23}(\mu_3,\mu_2,\mu_1)=
\mathrm{L}_{23}(\mu_3,\mu_2,\mu_1)
\mathrm{L}_{13}(\mu_1,\mu_2,\mu_3)
\mathrm{R}_{12}(\mu_1,\mu_2,\mu_3).
\EN

It turns out that the matrix structure of the Lax operator normalizing
its elements by the energy weight $c$ is,
\EQ
\label{LAXP}
\mathrm{L}(\mu_1,\mu_2,\mu_3)=\left(
\begin{array}{cccc}
\bar{a}(\mu_1,\mu_2,\mu_3) & 0 & 0 & 0 \\
0 & \bar{b}(\mu_1,\mu_2,\mu_3) & 1 & 0 \\
0 & 1 & \bar{b}(\mu_1,\mu_2,\mu_3) & 0 \\
0 & 0 & 0 & \bar{a}(\mu_1,\mu_2,\mu_3) \\
\end{array}
\right),
\EN
where the expressions for the rational functions 
$\bar{a}(\mu_1,\mu_2,\mu_3)$ and
$\bar{b}(\mu_1,\mu_2,\mu_3)$ are given by,
\EQ
\bar{a}(\mu_1,\mu_2,\mu_3)=\frac{\mu_2-\mu_1-\mu_3+\mu_1 \mu_3}{\mu_1-\mu_3-\mu_1 \mu_2 +\mu_1 \mu_3},~~
\bar{b}(\mu_1,\mu_2,\mu_3)=\frac{\mu_2(1-\mu_1)}{\mu_1-\mu_3-\mu_1 \mu_2 +\mu_1 \mu_3},\\
\label{LAXwei}
\EN

By the same token the form of the $\mathrm{R}$-matrix can be presented as,
\EQ
\label{RMAP}
\mathrm{R}(\mu_1,\mu_2,\mu_3)=\left(
\begin{array}{cccc}
\overline{\bf{a}}(\mu_1,\mu_2,\mu_3) & 0 & 0 & 0 \\
0 & \overline{\bf{b}}(\mu_1,\mu_2,\mu_3) & 1 & 0 \\
0 & 1 & \overline{\bf{b}}(\mu_1,\mu_2,\mu_3) & 0 \\
0 & 0 & 0 & \overline{\bf{a}}(\mu_1,\mu_2,\mu_3) \\
\end{array}
\right),
\EN
where the non-trivial entries can be computed with the help of Eqs.(\ref{ratios},\ref{LAXwei}). In this sense
we observe that Eq.(\ref{eqYB2}) tells us that the double prime energy weights are computed out of
Eq.(\ref{LAXwei}) by exchanging $\mu_1$ and $\mu_3$. Putting these information
together and after few simplifications we obtain,
\begin{eqnarray}
\label{RMAwei1}
&& \overline{\bf{a}}(\mu_1,\mu_2,\mu_3)=\frac{(\mu_1-\mu_3-\mu_1 \mu_3)(\mu_3-\mu_1+\mu_1 \mu_3-2\mu_2\mu_3+\mu_2^2)}{(\mu_1-\mu_3-\mu_1 \mu_2 +\mu_1 \mu_3)(\mu_1-\mu_3-\mu_1\mu_3+\mu_2\mu_3)},\\
\label{RMAwei2}
&& \overline{\bf{b}}(\mu_1,\mu_2,\mu_3)=\frac{\mu_2(\mu_3-\mu_1)(\mu_1-\mu_2+\mu_3 -\mu_1\mu_3)}{(\mu_1-\mu_3-\mu_1 \mu_2 +\mu_1 \mu_3)(\mu_1-\mu_3-\mu_1\mu_3+\mu_2\mu_3)}.
\end{eqnarray}

We finally show how the solution associated to the 
special divisor $\mathrm{Y}$ is obtained as 
particular case of Eqs.(\ref{eqYB2}-\ref{RMAwei2}). We just have to fix the spectral parameters by the
relations,
\EQ
\mu_1=\frac{(q-1)(t_1+t_2)}{(t_1+q)(t_2-1)},~
\mu_2=\frac{2q(t_1+t_2)}{(t_1+q)(t_2+q)},~
\mu_3=\frac{(q-1)(t_1+t_2)}{(t_1-1)(t_2+q)},
\EN
where $t_1$ and $t_2$ are free variables related to 
the affine parameterization of the
quadric (\ref{QUA}). Here we have also redefined $\Delta=q+1/q$. 

Considering this constraint in Eqs.(\ref{eqYB2}-\ref{RMAwei2}) we find that the
Yang-Baxter equation can be rewritten in the following additive form,
\EQ
\mathrm{L}_{12}(t_1/t_2)
\mathrm{L}_{13}(t_1)
\mathrm{L}_{23}(t_2)=
\mathrm{L}_{23}(t_2)
\mathrm{L}_{13}(t_1)
\mathrm{L}_{12}(t_1/t_2),
\EN
where Lax matrix assumes the standard form,
\EQ
\mathrm{L}(t)=\left(
\begin{array}{cccc}
\frac{q^2-t^2}{t(q^2-1)} & 0 & 0 & 0 \\
0 & \frac{q(1-t^2)}{t(q^2-1)} & 1 & 0 \\
0 & 1 & \frac{q(1-t^2)}{t(q^2-1)} & 0 \\
0& 0& 0& \frac{q^2-t^2}{t(q^2-1)}  \\
\end{array}
\right).
\EN

In the context of Algebraic Geometry such particular 
solution of the
Yang-Baxter equation corresponds 
to an affine algebraic group. In fact,
the $\mathrm{R}$-matrix elements are 
the generators of a map with the group law,
\EQ
\renewcommand{\arraystretch}{1.5}
\begin{array}{ccc}
\mathrm{D}(a^{'},b^{'},1) \times \mathrm{D}(a^{''},b^{''},1)&~~~{\longrightarrow}~~~ 
& \mathrm{D}({\bf{a}},{\bf{b}},1), 
\end{array}
\EN
since from Eq.(\ref{ratios}) we can  verify the identity 
${\bf{a}}^2+{\bf{b}}^2-1-\Delta {\bf{a}}{\bf{b}}=0$.

\section{Geometric Equivalences}
\label{sec3}

The Segre cubic threefold $\mathrm{S} \in \mathbb{CP}^4$ can be
defined by the vanishing of one homogeneous polynomial,
\EQ
\mathrm{S}=\{ (x_0:x_1:x_2:x_3:x_4) \in \mathbb{CP}^4 |x_0^3+x_1^3+x_2^3+x_3^3+x_4^3-(x_0+x_1+x_2+x_3+x_4)^3=0 \}.
\EN

It has been known since the end of the $19^{\mathrm{th}}$ century that the hypersurface $\mathrm{S}$ 
has rather special geometrical properties \cite{SEG}. This threefold has ten isolated singularities
of the simplest type $\mathrm{A}_1$ which is the maximum 
number permitted for a cubic hypersurface in $\mathbb{CP}^4$. This property characterizes 
such hypersurfaces
to be the Segre cubic $\mathrm{S}$ up to projective equivalences. The basic
properties of the Segre cubic primal and its relevance 
to modern research in Algebraic Geometry 
have been recently reviewed in \cite{DOL}. 

In the sequel we shall show that the
threefold $\mathrm{X}$ is birationally equivalent to the Segre 
cubic primal. Here we use the standard result in which the properties
of subsets of the product of projective spaces can be investigated
in the realm of the theory of plain projective 
varieties through the Segre embedding, see for example \cite{HAR}. 
In particular it follows that $\mathrm{X}$ can be realized as the inverse image
of some closed set of $\mathbb{CP}^8$. More specifically,
the Segre map takes a pair of points 
on $\mathbb{CP}^2 \times \mathbb{CP}^2$ to their pairwise products,
\EQ
\label{mapT1}
\renewcommand{\arraystretch}{1.5}
\begin{array}{ccc}
\mathbb{CP}^2 \times \mathbb{CP}^2 &~~~ \overset{\psi}{\longrightarrow}~~~ 
& \mathrm{Q}(z_{ij}) \subset \mathbb{CP}^8 \\
(a^{'}:b^{'}:c^{'})\times 
(a^{''}:b^{''}:c^{''}) 
& \longmapsto & (z_{00}:z_{01}:z_{02}:z_{10}:z_{11}:z_{12}:z_{20}:z_{21}:z_{22}),
\end{array}
\EN
where the coordinates are identified by the order $z_{00}=a^{'}a^{''},z_{01}=a^{'}b^{''}, 
\cdots,z_{22}=c^{'}c^{''}$ and the image $\mathrm{Q}(z_{ij})$ is the zero locus of
the quadratic equations, 
\EQ
\label{quad}
\mathrm{Q}(z_{ij})=z_{ij}z_{kl}-z_{il}z_{kj},~0 \leq i < k \leq 2~~\mathrm{and}~~0 \leq j <l \leq 2.
\EN

As the next step to define $\mathrm{X}$ as subvariety of $\mathbb{CP}^8$
we rewrite the bihomogenous polynomial (\ref{POLBI}) in the
new variables $z_{ij}$. This is done by paring the single and double 
prime energy weights according to the Segre map (\ref{mapT1}) and as a result we obtain
the following homogenous polynomial,
\EQ
\mathrm{F}(z_{ij})=z_{00}z_{10}+z_{01}z_{11}-z_{02}z_{12}-z_{00}z_{01}-z_{10}z_{11}+z_{20}z_{21}
\EN

The desired system of equations for $\mathrm{X} \in \mathbb{CP}^8$ are then obtained
by adjoining  $\mathrm{F}(z_{ij})$ to the quadratic polynomials (\ref{quad}). Thus $\mathrm{X}$
can be redefined as follows,
\EQ
\mathrm{X}=\{ (z_{00}:z_{01}:\cdots:z_{22}) \in \mathbb{CP}^8 |\mathrm{F}(z_{ij})=\mathrm{Q}(z_{ij})=0 \}.
\EN

In order to get some insights on the geometric properties of $\mathrm{X}$ we 
consider this variety on the affine chart $z_{00} \neq 0$. In this open subspace
we can use the quadrics $\mathrm{Q}(z_{ij})=0$ to eliminate four of the variables such as for
example the coordinates $z_{11},z_{12},z_{21}$ and $z_{22}$. After this elimination we 
end up with an affine threefold $\mathrm{T}$ whose 
projective closure is given by,
\EQ
\overline{\mathrm{T}}=\{ (z_{00}:z_{01}:z_{02}:z_{10}:z_{20}) \in \mathbb{CP}^4 |z_{01}(z_{00}^2+z_{10}^2-z_{20}^2)-z_{10}(z_{00}^2+z_{01}^2-z_{02}^2)=0 \}.
\EN

The study of the singular locus of $\overline{\mathrm{T}}$ can again be done with the help of
Singular algebra system \cite{SIN}. We conclude that they are constituted of
ten points of type $\mathrm{A}_1$ and therefore 
the threefolds $\overline{\mathrm{T}}$ and $\mathrm{S}$ should be projectively equivalent. However, for concrete results,
we have to exhibit an explicit example of such linear automorphism which involves a cumbersome 
combinatorial analysis among projective coordinates. We have been fortunate to find one such possible
projective transformation, namely 
\EQ
\label{PROJ}
\left[
\begin{array}{c}
x_0  \\
x_1  \\
x_2  \\
x_3  \\
x_4  \\
\end{array}
\right]=
\left(
\begin{array}{ccccc}
1 & 1 & 0& -1 & 0 \\
0 & -1 & 0& 0 & 1 \\
0 & -1 & 0& 0 & -1 \\
0 & 0 & -1& 1 & 0 \\
0 & 0 & 1& 1 & 0 \\
\end{array} 
\right)
\left[
\begin{array}{c}
z_{00}  \\
z_{01}  \\
z_{02}  \\
z_{10}  \\
z_{20}  \\
\end{array}
\right].
\EN

At this point we have gathered the basic ingredients to built up morphisms between
the therefolds $\mathrm{X}$ and the Segre cubic $\mathrm{S}$ . Combining  
the elimination procedure described above with the linear transformation (\ref{PROJ})
we are able to establish the following rational map,
\EQ
\label{mapB}
\renewcommand{\arraystretch}{1.5}
\begin{array}{ccc}
\mathrm{S} \subset \mathbb{CP}^4 &~~~ \overset{\phi}{\longrightarrow}~~~ 
& \mathrm{X} \subset \mathbb{CP}^8 \\
(x_0:x_1:x_2:x_3:x_4) 
& \longmapsto & (\phi_{0}:\phi_{1}:\phi_{2}:\phi_{3}:\frac{\phi_{4}}{\phi_0}
:\frac{\phi_{5}}{\phi_0}:\phi_{6}:\frac{\phi_{7}}{\phi_0}:\frac{\phi_{8}}{\phi_0}),
\end{array}
\EN
where the expressions of the polynomials $\phi_j$ are,
\begin{eqnarray}
&& \phi_{0}= 2x_0 + x_1 + x_2 + x_3 + x_4,~
\phi_{1}= -x_1 - x_2,~
\phi_{2}= x_4 - x_3,~ 
\phi_{3}= x_3 + x_4,~
\phi_{4}= -(x_1 + x_2)(x_3 + x_4) \nonumber \\
&& \phi_{5}= -(x_3 - x_4)(x_3 + x_4),~ 
\phi_{6}= x_1 - x_2,~
\phi_{7}= -(x_1 - x_2)(x_1 + x_2),~
\phi_{8}= -(x_1 - x_2)(x_3 - x_4).
\end{eqnarray}

This map is well defined since the closure of the 
image of $\phi$ is the whole of $\mathrm{X}$.
It is also invertible and the respective inverse can the written as,
\EQ
\renewcommand{\arraystretch}{1.5}
\begin{array}{ccc}
\mathrm{X} \subset \mathbb{CP}^8 &~~~ \overset{\phi^{-1}}{\longrightarrow}~~~ 
& \mathrm{S} \subset \mathbb{CP}^4 \\
(z_{00}:z_{01}:\cdots:z_{22}) 
& \longmapsto & (z_{00}+z_{01}-z_{10}:z_{20}-z_{01}:-z_{01}-z_{20}:z_{10}-z_{02}:z_{10}+z_{02}),
\end{array}
\EN
completing the proof that $\mathrm{X}$ and $\mathrm{S}$ are birational varieties.

\subsection{Parameterization}

In order to obtain a parameterization for the energy 
weights it is sufficient to establish
a one-to-one map from $\overline{\mathrm{T}}$ to $\mathbb{CP}^3$. 
This birational equivalence
can be set up exploring the generators of linear system of quadrics passing through
five generic points. The technical details of this construction for the Segre cubic primal
can, for instance, be found on the pages 149-152 of reference \cite{BAK}. 
This method can be adapted to 
$\overline{\mathrm{T}}$ and
in what follows
we present only the final result. 
Let $\lambda_0,\cdots,\lambda_3$ be
the coordinates of $\mathbb{CP}^3$ then the desired map is,
\EQ
\label{mapC}
\renewcommand{\arraystretch}{1.5}
\begin{array}{ccc}
\mathbb{CP}^3 &~~~ \overset{\varphi}{\longrightarrow}~~~ 
& \overline{\mathrm{T}} \subset \mathbb{CP}^4 \\
(\lambda_0:\lambda_1:\lambda_2:\lambda_3) 
& \longmapsto & (\varphi_{0}:\varphi_{1}:\varphi_{2}:\varphi_{3}:\varphi_{4}),
\end{array}
\EN
where the polynomials $\varphi_j$ are,
\begin{eqnarray}
&& \varphi_0 = -\lambda_0\lambda_2 + \lambda_3(\lambda_0 - \lambda_1 + \lambda_2),~
\varphi_1 = \lambda_1(\lambda_2 - \lambda_3),~
\varphi_3= \lambda_1(\lambda_0 - \lambda_3), \nonumber \\
&&\varphi_2= \lambda_2(\lambda_1 - \lambda_3) + \lambda_0(-\lambda_2 + \lambda_3),~
\varphi_4= -\lambda_2\lambda_3+ \lambda_0(-\lambda_1 + \lambda_2 + \lambda_3).
\end{eqnarray}

By taking into account the coordinates identification of 
the Segre embedding (\ref{mapT1}) we find
that the energy weights ratios are given by,
\begin{eqnarray}
&& \frac{a^{'}}{c^{'}}=\frac{\lambda_3(\lambda_0-\lambda_1+\lambda_2)-\lambda_0\lambda_2}
{\lambda_0(\lambda_2-\lambda_1+\lambda_3)-\lambda_2\lambda_3},~
\frac{b^{'}}{c^{'}}=\frac{\lambda_1(\lambda_0-\lambda_3)}
{\lambda_0(\lambda_2-\lambda_1+\lambda_3)-\lambda_2\lambda_3}, \nonumber \\
&& \frac{a^{''}}{c^{''}}=\frac{\lambda_3(\lambda_2-\lambda_1+\lambda_0)-\lambda_0\lambda_2}
{\lambda_2(\lambda_0-\lambda_1+\lambda_3)-\lambda_0\lambda_3},~
\frac{b^{''}}{c^{''}}=\frac{\lambda_1(\lambda_2-\lambda_3)}
{\lambda_2(\lambda_0-\lambda_1+\lambda_3)-\lambda_0\lambda_3}. \nonumber \\
\end{eqnarray}

We now note that the single and double prime ratios are related via the
exchange $\lambda_0 \leftrightarrow \lambda_2$.
After defining the affine parameters $\mu_1=\lambda_0/\lambda_3$,
$\mu_2=\lambda_1/\lambda_3$ 
and $\mu_3=\lambda_2/\lambda_3$ we are able to obtain
Eqs.(\ref{eqYB2},\ref{LAXwei}).

\section{Conclusions}
\label{CONCLU}
In this paper we have showed that the Baxter's integrability condition
for the symmetrical six-vertex model leads us to an algebraic variety
which is birationally isomorphic to the Segre cubic primal. This
threefold has unique geometric properties which are used to provide
a parameterization for the $\mathrm{R}$-matrix and Lax operators triple.
In general one expects that generic solutions of the Yang-Baxter
equation will involve the study of varieties defined as subsets of the
product of two projective spaces. It appears of interest to
investigate the geometrical properties of such varieties, with
the help of the Segre embedding, for other integrable models of
statistical mechanics. 
This study could reveal whether
the ubiquity found here is accidental
or may be also present in different ways for other classes of
statistical configurations.

\section*{Acknowledgments}
This work has been partially supported by the Brazilian Research Agencies 
FAPESP and CNPq. I am grateful to Daniel Levcovitz
and Eduardo Tengan for fruitful discussions on Algebraic Geometry topics 
over the past years.

\end{document}